\documentclass{aa}
\usepackage{psfig,latexsym,longtable,lscape}

\def\lsim{\lower0.5ex\hbox{$\; \buildrel < \over \sim \;$}}
\def\ros{{\sl ROSAT }}

\def\ergsec{\hbox{erg s$^{-1}$ }}

\def\it{\sl}
\def\asec{\ifmmode ^{\prime\prime}\else$^{\prime\prime}$\fi}
\def\amin{\ifmmode ^{\prime}\else$^{\prime}$\fi}

\begin{document}

   \thesaurus{6(02.01.2; 08.09.2; 08.13.1; 08.14.1; 08.18.1; 13.25.5)
             } 
   \title{The recent pulse period evolution of SMC~X-1}

   \author{P. Kahabka\inst{1} and X.-D. Li\inst{2}
          }

   \offprints{ptk@astro.uva.nl}
 
   \institute{$^1$~Astronomical Institute and Center for High Energy
   Astrophysics, University of Amsterdam, Kruislaan 403, NL-1098~SJ 
   Amsterdam, The Netherlands \\
   $^2$~Department of Astronomy, Nanjing University, Nanjing 210093,
   P.R. China (email:lixd@netra.nju.edu.cn)}

   \date{Received 23 December 1997 / Accepted 9 February 1999}
 
   \maketitle
   \markboth{P. Kahabka: Pulse Period Evolution of SMC~X-1}{}
 
   \begin{abstract}
%______________________________________ Do not leave a blank line here!
We report observations of SMC~X-1 in three new high-intensity states with  
{\sl ROSAT HRI} in December 1995, May 1997, and March 1998 in which pulsations
with a period of 0.70769 ($\pm$~0.00006)~s, 0.70706 ($\pm$~0.00001)~s, and
0.706707 ($\pm$~0.00001)~s respectively were detected. Combining the pulse 
periods
from observations with {\sl ROSAT PSPC} in high-intensity states in October 
1991 and March 1998, respectively, we obtain the spin-up rate of the pulsar in
recent 6.5~years, $-\dot{P}=(1.18\pm 0.06)\ 10^{-11}\ s\ s^{-1}$, consistent 
with the average spin-up rate $-\dot{P}=1.2\ 10^{-11}\ s\ s^{-1}$ determined 
from previous measurements indicating that the stable spin-up has continued.
Pulsations with a period of 0.709103 ($\pm$~0.000003) s were also detected 
$\sim$2~weeks after an X-ray turn-off during an X-ray low-intensity state 
in October 1991, and the period derivative derived within $\sim 10$ days is 
$-\dot{P}\simeq (1.1\pm 0.7)\ 10^{-11} s s^{-1}$. This is
consistent with a constant accretion torque in sign and magnitude. 
The magnitude of the magnetic moment of the pulsar is discussed based on
different description of the apparent spin-up behavior.

%  14.Sep.'90: Demo-Vs.
%_____________________________________ Do not leave a blank line here!
   \end{abstract}
 
%
%  14.Sep.'90: Demo-Vs.
%________________________________________________________________
 
\section{Introduction}

SMC~X-1 was detected during a rocket flight (Price et al. 1971). 
The discovery of eclipses with the {\it Uhuru} satellite with a period 
of 3.89~days (Schreier et al. 1972) established the binary nature of 
the source. The optical counterpart Sk~160 has been identified as a 
B0~I supergiant (Webster et al. 1972; Liller 1973). Optical photometry
indicated the presence of an accretion disk influencing the optical
light curve (van Paradijs \& Zuiderwijk 1977). In X-rays both low- and
high-intensity states have been observed with an X-ray luminosity $L_x$ 
varying from $\sim 10^{37} \ergsec$ to $\sim 5\times 10^{38} \ergsec$ 
(Schreier et al. 1972;
Tuohy \& Rapley 1975; Seward \& Mitchell 1981; Bonnet-Bidaud \& van der Klis 
1981). Angelini et al. (1991) discovered an X-ray burst from SMC~X-1 probably 
from type~II like in the Rapid Burster generated by an instability in the 
accretion flow. A $\sim 60$~day quasi-periodicity was suggested by Gruber \& 
Rothschild (1984) from {\it HEAO~1~(A4)} data, and was confirmed by more 
recent {\sl RXTE} observations (Levine et al. 1996; Wojdowski et al. 1998). 
A pulse period of $P=0.71$ sec (Lucke et al. 1976), neutron star mass 
$M_{\rm x}=0.8-1.8 M_{\odot}$, companion mass $M_{\rm c}\approx 19 M_{\odot}$ 
and companion radius $R_{\rm c}\approx 18 R_{\odot}$ (Primini, Rappaport \& 
Joss 1977) are well established. A decay in the orbital period $\dot{P}_{\rm 
orb}/P_{\rm orb} = (-3.36\pm0.02)\times 10^{-6} yr^{-1}$ was found (Levine et 
al. 1993), probably due to tidal interaction between the orbit and the 
rotation of the companion star, which is supposed to be in the hydrogen shell 
burning phase. Li \& van den Heuvel (1997) argued that the magnetic moment in 
SMC X-1 may be low like that of the bursting pulsar GRO~J1744-28, i.e. 
$\sim 10^{29}\ G\ cm^3$. 

%
%  14.Sep.'90: Demo-Vs.
%__________________________________________________________________
 
\section{Observations}

The observations reported in this paper have been performed with the
{\sl PSPC} and {\sl HRI} detectors of the {\sl ROSAT} satellite (Tr\"umper
1983). In Table~1 a log of the observations analysed in this work 
is given. The observations were centered on SMC~X-1. The October-1991 
observations and the June-1993 observation have been retrieved from the 
public \ros archive in November 1997. The \ros {\sl HRI} observations were 
made by the first author of this paper. The recently discovered transient 
RX~J0117.6-7330 located $\sim 5\amin $ southeast of SMC~X-1 (Clark, 
Remillard \& Woo 1996) has not been detected in these observations.

\subsection{High- and low-intensity X-ray states}

SMC~X-1 has been observed during the {\sl ROSAT} all-sky survey (Kahabka \& 
Pietsch 1996). The source  was in a low-intensity state in a first pointed 
observation, and was found in a high state in a {\sl ROSAT PSPC} pointing in 
October 1991, preceding the low-intensity state by $\sim$12~days. This
limits the duration of this specific X-ray turn-off phase to less than 
2~weeks. In this paper the pulse periods of SMC X-1 during three X-ray high 
states observed with the {\sl ROSAT HRI} $\sim$4, $\sim$5.5, and 
$\sim$6.5~years after the 1991 high state observation are reported. Pulse 
period determinations from the {\sl ROSAT} observations are summarized 
in Table~1.

\subsection{Pulse periods and period derivatives}

We have searched for the pulse period in the data from four high state data 
and from the low state data following the first high state. In the present 
analysis the event times have been projected from the spacecraft to the 
solar-system barycenter with standard {\it EXSAS} software employed 
(Zimmermann et al. 1994). They have also been corrected for arrival time 
delays in the binary orbit by use of the ephemeris and the orbital solution 
given in Levine et al. (1993). This takes into account the change in the 
length of the orbital period and of the mid eclipse ephemeris due to orbital 
decay (Wojdowski et al. 1998). Period uncertainties have been determined from 
the relation $\delta P = P/(T_{obs}\times N_{bin})$, with the exposure time 
$T_{obs}$ given in Tab~1 and the number of phase bins $N_{\rm bin}=10$. 

Periods of $P$=0.709113 ($\pm$~0.000003) s ($\chi^2$=3000, 9 degrees of 
freedom), $P$=0.708600 ($\pm$ 0.000002) s ($\chi^2$=980, 9 degrees of 
freedom), P=0.70769 ($\pm$~0.00006) s ($\chi^2$=56, 9 degrees of freedom), 
$P$=0.707065 ($\pm$ 0.000010) s ($\chi^2$=113, 9 degrees of freedom) and 
$P$=0.70670 ($\pm$ 0.00002) s ($\chi^2$=250, 9 degrees of freedom) have been 
obtained during the October-1991, June-1993, December-1995, May-1997 and the March-1998 
high-intensity states, respectively (cf. Figure~1 and Table~1). From the 
October-1991 to the December-1995 high state a change in pulse period with a mean 
$-\dot{P}=(1.08\pm 0.05)\ 10^{-11}\ s\ s^{-1}$ and from the June-1993 to the 
March-1998  high state a mean $-\dot{P}= (1.25\pm 0.08)\ 10^{-11}\ s\ s^{-1}$ 
are derived. 

The period derivative derived over the $\sim$6~year interval from October-1991 
to Mar-1998 is $-\dot{P}=(1.18\pm 0.06) \ 10^{-11}\ s\ s^{-1}$ consistent with 
the mean $-\dot{P}=1.20\ 10^{-11} \ s\ s^{-1}$ derived from previous 
observations (Levine et al. 1993). The evolution of the pulse period with 
Julian date using data from Henry \& Schreier (1977), Kunz et al. (1993), 
Levine et al. (1993), Wojdowski et al. (1998) and the results from this work 
is given in Figure~2. Also shown are the residuals compared to a linear 
best-fit with a $-\dot{P} = 1.153\ 10^{-11}\ s s^{-1}$. It is very evident 
that the pulse period of SMC~X-1 undergoes a period walk with a time scale of 
a few 1000~days (a few years). But the amplitude of this period walk is small
($\sim 1.5\ 10^{-4}\ s$). It may be suspected that somewhere at the end
of 1994 the ``positive'' deviation from the mean $\dot{P}$ was largest
(cf Figure~2). After this time the mean $\dot{P}$ may have increased. It is
not clear in which way the period walk continues. An explanation of this
period walk in terms of a ``free'' precessing neutron star is unlikely
(cf. Bisnovatyi-Kogan \& Kahabka 1993).

%                                     Two column figure (place early!)
%______________________________________________ Gamma_1 (lg rho, lg e)
   \begin{figure}
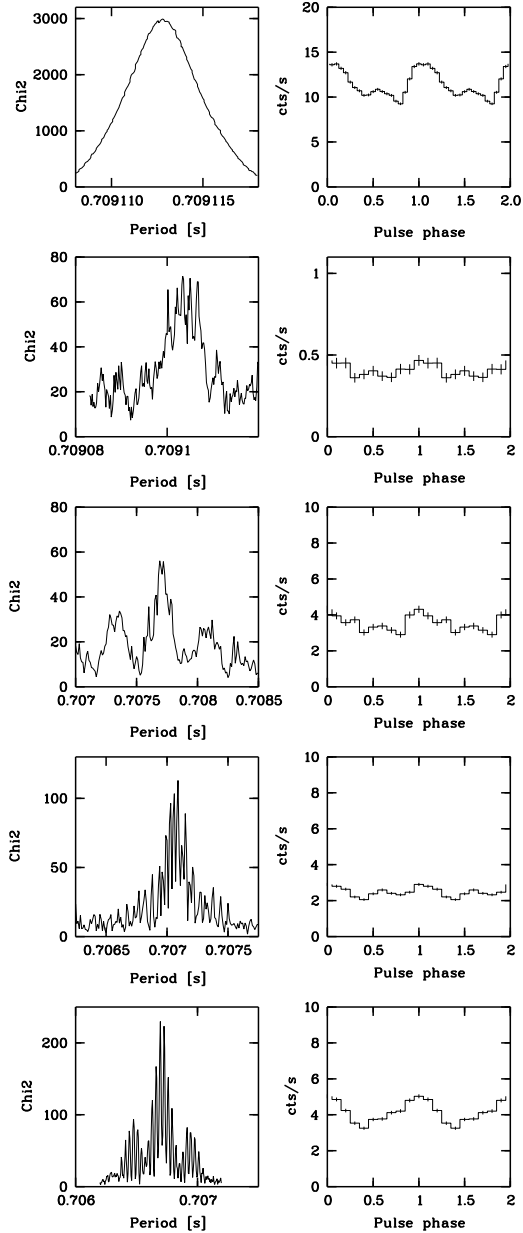

      \centering{
      \vbox{\psfig{figure=8532.f1,width=7.0cm,%
       bbllx=1.5cm,bblly=1.7cm,bburx=13.0cm,bbury=7.2cm,clip=}}\par
      \vbox{\psfig{figure=8532.f2,width=7.0cm,%
       bbllx=1.5cm,bblly=1.7cm,bburx=13.0cm,bbury=7.2cm,clip=}}\par
      \vbox{\psfig{figure=8532.f3,width=7.0cm,%
       bbllx=1.5cm,bblly=1.7cm,bburx=13.0cm,bbury=7.2cm,clip=}}\par
      \vbox{\psfig{figure=8532.f4,width=7.0cm,%
       bbllx=1.5cm,bblly=1.7cm,bburx=13.0cm,bbury=7.2cm,clip=}}\par
      \vbox{\psfig{figure=8532.f5,width=7.0cm,%
       bbllx=1.5cm,bblly=1.7cm,bburx=13.0cm,bbury=7.2cm,clip=}}\par
       }
      \caption[]{$\chi^2$ distribution and pulse profile for period search 
                 applied to period data of SMC~X-1 in Oct-1991 (high, 
                 low-state), Dec-1995, May-1996 and Mar-1998 (from top to 
                 bottom). Pulse phase 1.0 referes to the pulse maximum.}
         \label{FigGam}
    \end{figure}

A pulse period search has also been performed in an observation during a 
low-intensity state performed in the time interval 16 to 19 Oct-1991 
(cf. Table~1 and Figure~1). A period of $P$=0.709103 $\pm$~0.000003~s 
($\chi^2$=71, 9 degrees of freedom) has been determined (cf. Figure~1). 
This period is close to the period determined during the 7-Oct to 8-Oct-1991 
high-intensity state and consistent with the long-term negative $\dot{P}$ 
value. The significance of this period is $8\sigma$. The period derivative 
between the high and low state in Oct-1991 (with a time interval $\sim 10$ 
days) is $-\dot{P}\simeq (0.4-1.8)\ 10^{-11} s\,s^{-1}$.

%                                     Two column figure (place early!)
%______________________________________________ Gamma_1 (lg rho, lg e)
   \begin{figure}
      \centering{
      \vbox{\psfig{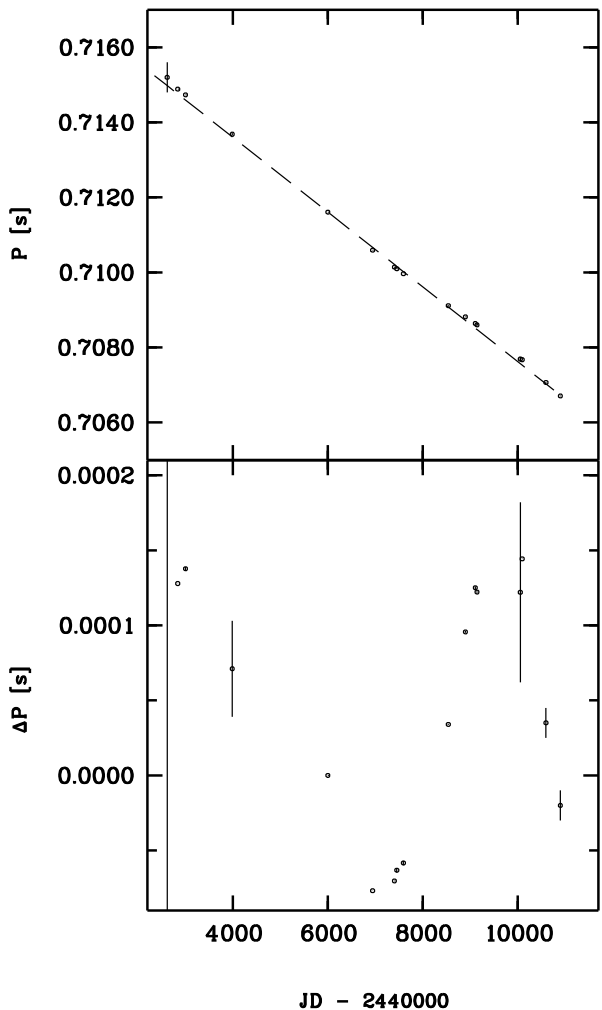}}\par
       }
      \caption[]{Upper panel: 22.7~year pulse period history of SMC~X-1 
                 as a function of Julian date. Values are shown for 
                 observations with {\sl Apollo-Soyuz}, {\sl SAS-3}, 
                 {\sl Ariel V}, {\sl Einstein}, {\sl EXOSAT}, {\sl Ginga}, 
                 {\sl HEXE}, {\sl ROSAT}, {\sl ASCA}, and {\sl RXTE} 
                 (cf. Table~2 for a summary). The best-fit mean $-\dot{P} 
                 = 1.153\ 10^{-11}\ s s^{-1}$ is given as dashed line. 
                 Lower panel: Residuals of least-square linear fit to the 
                 period values.}
         \label{FigGam}
    \end{figure}

    \begin{table*}
      \caption[]{{\sl ROSAT} observations of SMC X-1}
      \begin{flushleft}
      \begin{tabular}{lcccccc}
      \hline
      \hline
      \noalign{\smallskip}
 Instr.   &ID       &\multicolumn{2}{c}{Time start - end}                  &Expos.&Pulse period          &State\\
          &         &       (UT)                       &  (JD - 2440000)   &(ksec)&      (s)             &         \\
      \noalign{\smallskip}
      \hline
      \noalign{\smallskip}
{\sl PSPC}&400022p  & 7-Oct-1991 04:09 - 8-Oct-1991 02:41 & 8536.673 - 8537.612&16.6  &0.709113$\pm$0.000003 & High\\
{\sl PSPC}&600197p  &16-Oct-1991 07:34 - 19-Oct-1991 23:32& 8545.815 - 8549.481&22.1  &0.709103$\pm$0.000003 & Low \\
{\sl PSPC}&400022p-2& 3-Jun-1993 00:03 - 4-Jun-1993 04:40 & 9141.502 - 9142.694&11.8  &0.708600$\pm$0.000002 & High\\
{\sl HRI} &500420h-2& 2-Dec-1995 23:09 - 2-Dec-1995 23:39 &10054.465 -10054.486& 1.08 &0.70769$\pm$0.00006   & High\\
{\sl HRI} &500420h-4&15-May-1997 13:36 - 4-Jun-1997 22:29 &10584.067 -10604.437& 4.89 &0.707065$\pm$0.000010 & High\\
{\sl HRI} &500420h-5&25-Mar-1998 17:03 - 26-Mar-1998 17:09&10898.210 -10899.214& 4.24 &0.706707$\pm$0.000010 & High\\
      \noalign{\smallskip}
      \hline
      \end{tabular}
      \end{flushleft}
   \end{table*}

    \begin{table}
      \caption[]{Periods and period residuals from least-square linear fit 
                 for observations of SMC X-1. The periods derived from the
                 {\sl Einstein} and {\sl HEXE} observations have not been
                 corrected for orbital decay.}
      \begin{flushleft}
      \begin{tabular}{lrllc}
      \hline
      \hline
      \noalign{\smallskip}
 Mission           & JD        & Pulse period     & Residual   & Ref. \\
                   & - 2440000 &     [s]          &      [s]   &      \\
      \noalign{\smallskip}
      \hline
      \noalign{\smallskip}
{\sl Uhuru}        &  1114.7   & 0.71748(26)     &  +0.00100   & [1]  \\
{\sl Apollo}       &  2614.3   & 0.7152(4)       &  +0.00022   & [1]  \\
{\sl -Soyuz}       &           &                 &             &      \\
{\sl SAS 3}        &  2836.7   & 0.71488585(4)   &  +0.000128  & [2]  \\
{\sl Ariel V}      &  3000.2   & 0.7147337(12)   &  +0.000138  & [2]  \\
{\sl Einstein}     &  3986.4   & 0.713684(32)    &  +0.0000710 & [2]  \\
{\sl EXOSAT}       &  5999.0   & 0.7116079958(4) &  +0.000000  & [3]  \\
{\sl Ginga}        &  6942.0   & 0.710592116(36) & --0.0000769 & [2,4]\\
{\sl Ginga}        &  7401.2   & 0.710140670(15) & --0.0000703 & [2,4]\\
{\sl HEXE}         &  7452.0   & 0.7100978(15)   & --0.0000632 & [3]  \\
{\sl HEXE}         &  7591.5   & 0.7099636(15)   & --0.0000584 & [3]  \\
{\sl ROSAT}        &  8537.1   & 0.70911393(34)  &  +0.0000340 & [2]  \\
{\sl ROSAT}        &  8897.2   & 0.7088166(6)    &  +0.0000956 & [2]  \\
{\sl ASCA}         &  9105.0   & 0.7086390(9)    &  +0.000125  & [2]  \\
{\sl ROSAT}        &  9142.0   & 0.7085992(5)    &  +0.000122  & [2]  \\
{\sl ROSAT}        & 10054.5   & 0.70769(6)      &  +0.000122  & [5]  \\
{\sl RXTE}         & 10093.5   & 0.70767330(8)   &  +0.000144  & [2]  \\
{\sl ROSAT}        & 10594.3   & 0.707065(10)    &  +0.0000350 & [5]  \\   
{\sl ROSAT}        & 10898.7   & 0.706707(10)    & --0.0000200 & [5]  \\
      \noalign{\smallskip}
      \hline
      \end{tabular}
      \end{flushleft}
Ref.: [1] Henry \& Schreier 1977; [2] Wojdowski et al. 1998; 
[3] Kunz et al. 1993; [4] Levine et al. 1993; [5] this work.
   \end{table}

\section{Discussion}

Disk-fed magnetic neutron stars have been predicted to experience spin-up 
or spin-down episodes due to the torque exerted by the accretion disk. 
In the magnetically-threaded accretion disk model, first suggested by 
Ghosh \& Lamb (1979a, 1979b), the spin-up rate $-\dot{P}$ is given by
\begin{equation}
-2\pi I\dot{P}/P^2 = \dot{M} (GM r_0)^{1/2} n (\omega_s),
\end{equation}
where $I$ is the moment of inertia of the neutron star, $\dot{M}$ the 
accretion rate, $G$ the gravity constant and $r_0$ the inner edge of the 
accretion disk. The dimensionless torque $n(\omega_s)$, which includes
the torque contribution from both matter accretion and magnetic stress,
is a function of the ``fastness parameter" 
$\omega_s\equiv (r_0/r_c)^{3/2}$, where $r_c\equiv 
(GMP^2/4\pi^2)^{1/3}$ is the corotation radius. 

With the {\sl mean} spin-up rate and X-ray luminosity observed in SMC X-1, 
equation (1) has two sets of solutions: (1) the magnetic moment of the
pulsar $\mu$ is less than a few $10^{29} G\,cm^3$ with $r_0\ll r_c$;
(2) $\mu$ is around $10^{30} G\,cm^3$ with $r_0\sim r_c$
(Li \& van den Heuvel 1997). If the X-ray intensity during the low state, 
which is lower than that during the high state by a factor of $35-50$, 
is due to a reduction in the mass accretion rate, the condition $r_0\le r_c$ 
implies $\mu\lsim 2 \times 10^{29} G\,cm^3$. However, Gruber \& 
Rothschild (1984; see also Levine et al. 1996; Wojdowski et al. 1998) 
have suggested
the possibility of modulation of the observed X-ray intensity by a tilted,
precessing, accretion disk like that in Her X-1 (Katz 1973), which would
imply that the intrinsic X-ray luminosity or mass accretion rate of the 
pulsar could be quite steady. In this case the pulsar magnetic moment may 
lie between $10^{29} G\, cm^3$ and $10^{30} G\,cm^3$, though a magnetic 
moment as high as $\sim 10^{30} G\,cm^3$ seems less likely for SMC X-1, 
because of the following arguments. As seen in Fig. 2, the spin-up rate of 
the pulsar in 1980s and 1990s varied around its mean value by $\sim 20\%$. 
The most straightforward explanation for this change is that the accretion 
rate has fluctuated by a similar (or a bit larger) factor\footnote{Due to 
disk precessing, the accretion torque also changes, but on a timescale much 
shorter than that of the pulse period variation.}. If $\mu\sim 10^{30} 
G\,cm^3$, $r_0$ would be so close to $r_c$ that the pulsar would spin down 
when the accretion rate decreased by a small factor (less than 20\%), 
contradicted with the steady spin-up observed.

The above arguments are based on the classical accretion torque models,
which, however, has encountered difficulties in explaining the period
evolution in the X-ray pulsar Cen X-3, which possesses many similarities 
with SMC X-1.  Both pulsars are in a close binary system 
with a supergiant companion star overflowing its Roche-lobe, 
accreting from a disk (Tjemkes et al. 1986) at a high rate, 
with the X-ray luminosities close to or higher than the Eddington luminosity 
(Nagase 1989).  Tidal torque between the distorted supergiant and the 
neutron star leads to an orbital decay at a similar rate in the two systems
(cf. White et al. 1995). 
However, Cen X-3 exhibits a pulse period evolution quite different from 
SMC X-1. Prior to 1991, Cen X-3 had already been found to show a secular 
slow spin-up superposed with fluctuations and 
short episodes of spin-down. The more frequently sampled {\sl BATSE} data
show that Cen X-3 exhibits 10-100 d intervals of steady spin-up and spin-down
at a much larger rate, and the long-term ($\sim$ years) spin-up 
trend is actually the average consequence of the frequent 
transitions between spin-up and spin-down (Finger et al. 1994). Such spin 
behavior has been found in at least 4 out of 8 persistent X-ray pulsars
observed with {\sl BATSE} (Bildsten et al. 1997), and is difficult
to explain in terms of classical accretion torque models, which would require 
finely tuned step-function-like changes in the mass accretion rate.

It is interesting to see whether the secular spin-up in SMC X-1 
actually consists of transitions of short-term spin-up and spin-down, 
as in Cen X-3. If it is the case, this would indicate a larger instantaneous
accretion torque, and hence a higher magnetic moment. 

\section{Summary}

New values for the spin period of SMC~X-1 have been determined during 
recent \ros observations. These observations clearly show that the 
steady spin-up of the neutron star has continued. This makes SMC X-1 the
exceptional X-ray pulsar in which no spin-down episode has been observed,
though it is still  unknown whether this spin-up occurs in reality or
it is just an apparent superposition of more frequent spin-up and spin-down
episodes. The magnetic moment of the pulsar could be as small as $\sim
10^{29} G\,cm^3$ if its spin trend can be described by classical accretion
torque models. Detailed timing observations are strongly recommanded to
resolve this subject, and will have important implications on the theoretical
models for angular momentum transfer between a magnetic neutron star and
the surrounding accretion disk.

\acknowledgements
P.K. thanks H.Henrichs and P.Ghosh for helpful discussions. I thank E.P.J. 
van den Heuvel for reading the article. This research was supported in part 
by the Netherlands Organisation for Scientific Research (NWO) through Spinoza
Grant 08-0 to E.P.J. van den Heuvel. The {\sl ROSAT} project is supported
by the Max-Planck-Gesellschaft and the Bundesministerium f\"ur Forschung
und Technologie (BMFT). 

%________________________________________ Do not leave a blank line here!

%  14.Sep.'90: Demo-Vs.
%_____________________________________________________________________

\end{document}